\documentstyle[sprocl]{article}

\input epsf

 \newcommand{\zr}[1]{\mbox{\hspace*{#1em}}}
 \newcommand{\ID}{\mbox{{\sf 1}\zr{-0.14}\rule{0.04em}{1.55ex}\zr{0.1}}}

\begin{document}

{\noindent\small FAU-TP3-00/10 \hfill  UNITU-THEP-00/15 
\hfill hep-ph/0009219}\vspace{5mm} 

\title{What the Infrared Behavior of QCD Green Functions can tell
       us about Confinement in the Covariant Gauge\footnote{Invited talk
       by L. v. Smekal at ``Quark Confinement and the Hadron
       Spectrum IV'', Vienna, July 4-8, 2000.} 
}

\author{Lorenz von Smekal}
\address{Universit\"at Erlangen-N\"urnberg, 
	 Institut f\"ur Theoretische Physik III,\\
	 Staudtstr.~7, 91058 Erlangen, Germany\\
	 E-mail: smekal@theorie3.physik.uni-erlangen.de}

\author{Reinhard Alkofer}
\address{Universit\"at T\"ubingen, Institut f\"ur Theoretische Physik,\\
         Auf der Morgenstelle 14, 72076 T\"ubingen, Germany\\
	 E-mail: Reinhard.Alkofer@uni-tuebingen.de}

\maketitle	 
	 

\begin{abstract}
We review aspects of confinement in the covariant and local 
description of QCD and discuss to what extend our present knowledge of the
infrared behavior of QCD Green functions can support this description.
In particular, we emphasize: the 
positivity violations of transverse gluon and quark states, 
the Kugo-Ojima confinement criterion, and the conditions necessary to avoid 
the decomposition property for colored clusters.
We summarize how these issues relate to the infrared behavior of the 
propagators in Landau gauge QCD as extracted from solutions to truncated
Dyson-Schwinger equations and lattice simulations.  
\end{abstract}

\section{The Covariant Description of Confinement}
Covariant quantum theories of gauge fields require indefinite metric
spaces. Abandoning the positivity of the representation space
already implies some modifications to the standard
(axiomatic) framework of quantum field theory. 
Maintaining the much stronger principle of locality,
great emphasis has been put on the idea of relating confinement to the
violation of positivity in QCD.   
Just as in QED, where the Gupta-Bleuler prescription is to
enforce the Lorentz condition on physical states, a semi-definite {\em
physical subspace} can be defined as the kernel of an operator.  
The physical states then correspond to equivalence classes of states 
in this subspace differing by zero norm components.  
Besides transverse photons covariance implies the existence of 
longitudinal and scalar photons in QED. The latter two form metric partners
in the indefinite space. The Lorentz condition eliminates
half of these leaving unpaired states of zero norm which do not contribute to
observables. Since the Lorentz condition commutes with the 
$S$-Matrix, physical states scatter into physical ones exclusively. 

Confinement in QCD can be ascribed to an analogous
mechanism:
Within the framework of BRS algebra, in the simplest version for
the BRS-charge $Q_B$ and the ghost number $Q_c$ (both hermitean with respect
to an indefinite inner product) given by,
\begin{equation} 
           Q_B^2 = 0 \; , \quad \left[ iQ_c , Q_B \right] = Q_B \; ,
\end{equation}
completeness of the nilpotent BRS-charge $Q_B$ in a state space $\mathcal{V}$
of indefinite metric is assumed. This charge generates the BRS 
transformations by ghost number graded commutators $\{ \, , \}$,
{\it i.e.}, by commutators or anticommutators for fields with even
or odd ghost number, respectively.
The semi-definite {\em physical} subspace 
${\mathcal{V}}_{\mbox{\tiny phys}}  = \mbox{Ker}\, Q_B  $
is defined on the basis of this algebra by those states which are annihilated
by the BRS charge $Q_B$. 
Since $Q_B^2 =0 $, this subspace contains the space $ \mbox{Im}\, Q_B $
of so-called daughter
states which are images of others, their parent states in $\mathcal{V}$.
A physical Hilbert space is then obtained as (the completion of) the 
covariant space of equivalence classes, the BRS-cohomology of states in the
kernel modulo those in the image of $Q_B$,
\begin{equation}
     {\mathcal{H}}(Q_B,{\mathcal{V}}) = {\mbox{Ker}\, Q_B}/{\mbox{Im}\, Q_B} 
       \simeq  {\mathcal{V}}_s \; , 
\end{equation}
which is isomorphic to the space ${\mathcal{V}}_s$ of BRS singlets. 
It is easy to see that the image is furthermore contained in the orthogonal
complement of the kernel (given completeness they are identical).
It follows that states in $\mbox{Im}\, Q_B$
do not contribute to the inner product in $\mathcal{V_{\mbox{\tiny phys}}}$.   
Completeness is thereby important in the proof of positivity for physical
states,\cite{Kug79,Nak90} because it assures the absence of metric
partners of BRS-singlets, so-called ``singlet pairs''. 

With completeness, all states in $\mathcal{V}$ are either BRS
singlets in ${\mathcal{V}}_s$ or belong to so-called quartets which are 
metric-partner pairs of BRS-doublets (of parent with daughter states).
This exhausts all possibilities. The generalization of the
Gupta-Bleuler condition on physical states, $Q_B |\psi\rangle = 0$ in
$\mathcal{V}_{\mbox{\tiny phys}}$, eliminates half of these metric partners
leaving unpaired states of zero norm 
which do not contribute to any observable.
This essentially is the quartet mechanism: 
Just as in QED, one such quartet, the elementary quartet, is formed by
the massless asymptotic states of longitudinal and timelike gluons together 
with ghosts and antighosts which are thus all unobservable. 
In contrast to QED, however, one expects the quartet mechanism also 
to apply to transverse gluon and quark states, as far as they exist
asymptotically. A violation of positivity for such states then entails
that they have to be unobservable also. 
The combined evidence for
this collected below provides strong indication 
in favor of such a violation for possible transverse gluon states. 

The members of quartets are frequently said to be confined
kinematically. This is no comprehensive explanation of confinement but it is
one aspect in its present description.   
In particular, asymptotic transverse gluon and quark states
may or may not exist in the indefinite metric space $\mathcal{V}$. If either 
of them do exist and the Kugo-Ojima criterion is realized (see below), they
belong to unobservable quartets. In that case, the BRS-transformations of their
asymptotic fields entail that they form these quartets together with
ghost-gluon and/or ghost-quark bound states, respectively.\cite{Nak90}
It is furthermore  crucial for confinement, however, to have a mass gap in
transverse gluon correlations, {\it i.e.}, the massless transverse gluon
states of perturbation theory have to dissappear even though they should
belong to quartets due to color antiscreening and superconvergence in QCD for
less than 10 quark flavors.\cite{Oeh80,Nis94,Alk00} 

The quantum mechanical interpretation 
in terms of transition probabilities 
and measurements holds between physical states  and 
for expectation values in $\mathcal{V}_{\mbox{\tiny 
phys}}$ of those operators for which all contributions from zero norm
states vanish. For a (smeared local) operator $A$ to be
observable in this sense it is necessary to be BRS-closed, $\{ iQ_B , A \} \,
= 0 $, which coincides with the requirement of its local gauge invariance.
It then follows that for all states generated from the vacuum 
$|\Omega\rangle$ by any such observable one has
     $ \langle \Omega | A^\dagger A |\Omega \rangle \ge 0 $ .

If this construction is shown to apply to a QCD description of hadrons as the
genuine physical particles of $\mathcal{H}$, one concludes from physical
$S$-matrix unitarity (with respect to the indefinite inner product) that
absorptive thresholds in hadronic amplitudes can only be due to intermediate
hadronic states. The $S$-matrix commutes with the BRS-charge, {\it i.e.}, 
it is an observable in the above sense which transforms physical states into
physical ones exclusively and without measurable effects from
possible zero norm components.\cite{Nak90} One shows likewise in this
description that anomalous thresholds arise only from the
substructure of a given hadron as being composed of other hadrons. 
The argument to establish this employs standard analyticity 
properties for hadronic amplitudes and crossing to relate them to
absorptive singularities of other hadronic amplitudes which can in turn 
be due to intermediate hadronic states only.\cite{Oeh95}
 
Confinement depends on the 
realization of the unfixed global gauge symmetries in this formulation.
The identification of the 
BRS-singlets in the physical Hilbert space $\mathcal{H}$ with
color singlets is possible only if the charge of global gauge transformations
is BRS-exact {\em and} unbroken, {\it i.e.}, well-defined in the whole of the
indefinite metric space $\mathcal{V}$. The sufficent conditions for this are
provided by the Kugo-Ojima criterion: Considering the 
globally conserved current     
\begin{equation} 
    J^a_\mu = \partial_\nu F_{\mu\nu}^a  + \{ Q_B , D_{\mu}^{ab} \bar c^b \} 
    \qquad (\mbox{with} \; \partial_\mu J^a_\mu = 0 \,) \; ,
       \label{globG}
\end{equation}
one realizes that the first of its two terms corresponds to a coboundary 
with respect to the space-time exterior derivative while the second term 
is a BRS-coboundary with charges denoted by $G^a$ and $N^a$, respectively, 

\begin{equation} 
      Q^a =  \int d^3x \,  \partial_i F_{0 i}^a \,  +\,  \int d^3x \, 
             \{ Q_B , D_{0}^{ab} \bar c^b \} \, = \, G^a \, + \, N^a \; .
        \label{globC}
\end{equation}
For the first term herein there are only two options, it is either ill-defined
due to massless states in the spectrum of $\partial_\nu F_{\mu\nu}^a $, or else
it vanishes. 

In QED massless photon states contribute to the analogues of both currents
in~(\ref{globG}), and both charges on the r.h.s. in (\ref{globC}) are
separately ill-defined. One can employ an arbitrariness in the 
definition of the generator of the global gauge transformations
(\ref{globC}), however, to multiply the first term by a suitable
constant so chosen that these massless contributions cancel.
This way one obtains a well-defined and unbroken global
gauge charge which replaces the naive definition in (\ref{globC})
above.\cite{Kug95} Roughly speaking, there are two independent structures in
the globally conserved gauge currents in QED which both contain massless
photon contributions. These can be combined 
to yield one well-defined charge as the generator of global gauge
transformations leaving any other combination spontaneously broken,
such as the displacement symmetry which lead to the identification 
the photon with its massless Goldstone boson.\cite{Nak90,Fer71} 

If $\partial_\nu F_{\mu\nu}^a $ contains no massless
discrete spectrum on the other hand, {\it i.e.}, if there is no massless
particle pole in the Fourier transform of transverse gluon correlations, then
$G^a \equiv 0$.
In particular, this is the case for channels containing massive vector fields
in theories with Higgs mechanism, and it is expected to be also the case in
any color channel for QCD with confinement for which it actually represents one
of the two conditions formulated by Kugo and Ojima. 
In both these situations one first has equally, however, 
\begin{equation}
                       Q^a \, = \, N^a \, = \, \Big\{   Q_B \, , 
       \int d^3x \,   D_{0}^{ab} \bar c^b \Big\} \; ,
\end{equation}
which is BRS-exact. The second of the two conditions for confinement
is that this charge be well-defined in the whole of the indefinite metric space
$\mathcal{V}$. Together these conditions 
are sufficient to establish that all BRS-singlet physical
states in $\mathcal{H}$ are also color singlets, and that all colored states
are thus subject to the quartet mechanism. The 
second condition thereby provides the essential 
difference between Higgs mechanism and confinement. 
The operator $D_\mu^{ab}\bar c^b$ determining the charge $N^a$ will in
general contain a  {\em massless} contribution from the elementary
quartet due to the asymptotic field $\bar\gamma^a(x)$ in the  
antighost field,  $\bar c^a\, \stackrel{x_0 \to \pm\infty}{\longrightarrow}
\, \bar\gamma^a + \cdots $ (in the weak asymptotic limit), 
\begin{equation}
          D_\mu^{ab}\bar c^b \; \stackrel{x_0 \to \pm\infty}{\longrightarrow}
              \;   ( \delta^{ab} + u^{ab} )\,   \partial_\mu \bar\gamma^b(x) +
                 \cdots  \;  .
\end{equation}
Here, the dynamical parameters $ u^{ab} $ determine the contribution 
of the massless asymptotic state to the composite field $g f^{abc} A^c_\mu
\bar c^b  \, \stackrel{x_0 \to \pm\infty}{\longrightarrow}  \,
u^{ab} \partial_\mu \bar\gamma^b + \cdots $. These parameters can be obtained
in the limit $p^2\to 0$ from the Euclidean correlation functions of this
composite field, {\it e.g.},
\begin{equation}
\int d^4x \; e^{ip(x-y)} \,
\langle  \; D^{ae}_\mu c^e(x) \; gf^{bcd}A_\nu^d(y) \bar c^c (y) \; \rangle
\; =: \; \Big(\delta_{\mu \nu} -{p_\mu p_\nu \over p^2} \Big) \, u^{ab}(p^2)
\; .  \label{Corru}
\end{equation}
The theorem by Kugo and Ojima asserts that all $Q^a = N^a$ are
well-defined in the whole of  $\mathcal{V}$ (and do not suffer from
spontaneous breakdown), if and only if
\begin{eqnarray}
                 u^{ab} \equiv u^{ab}(0)  \stackrel{!}{=} - \delta^{ab} \; .
\label{KO1}
\end{eqnarray}
Then the massless states from the elementary quartet do not contribute to 
the spectrum of the current in $N^a$, and the equivalence between physical
BRS-singlet states and color singlets is established.\cite{Kug79,Nak90,Kug95}

In contrast, if $\mbox{det}(  \ID + u ) \not=0$, the global
gauge symmetry generated by the charges $Q^a$ in eq.~(\ref{globC}) is
spontaneuosly broken in each channel in which the gauge potential 
contains an asymptotic massive vector field.\cite{Kug79,Nak90} 
While this massive vector state 
results to be a BRS-singlet, the massless Goldstone boson states which 
usually occur in some components of the Higgs field, replace the 
third component of the vector field in the elementary
quartet and are thus unphysical. 
Since the broken charges
are BRS-exact, this {\em hidden} 
symmetry breaking is not directly observable in the Hilbert space of physical
states $\mathcal{H}$.  

It is nevertheless instructive to classify the different
szenarios according to the realization of the global gauge symmetry on the
whole of the indefinite metric space $\mathcal{V}$ of covariant gauge
theories. If it is unbroken, {\it i.e.}, as for QED and QCD, 
the first condition is crucial for confinement.
Namely, it is then necessary  to
have a mass gap in transverse gluon correlations ({\it i.e.}, in
$\partial_\nu F_{\mu\nu}^a $), since otherwise one could in principle have
{\em non-local} physical (BRS-singlet and thus gauge invariant) states which
are no color singlets, just as one has non-local gauge invariant
charged states in QED ({\it e.g.}, the state of one electron alone in the
world with its long-range Coulomb tail). 
Indeed, with unbroken global gauge invariance QED and QCD have in common
that any gauge invariant localized state must be
chargeless/colorless.\cite{Nak90} The
question is the extension to non-local states as approximated by local ones.
In QED this leads to the so-called charge superselection sectors,\cite{Haa96}
and non-local physical states with charge arise.    
If in QCD, with  unbroken global gauge symmetry {\em and}  mass gap, {\em
every} gauge-invariant state can be approximated by gauge-invariant localized
ones (which are colorless), one concludes that {\em every} gauge invariant
(BRS-singlet) state in $\mathcal{H}$ must be a color singlet.  

The (2nd condition in the) Kugo-Ojima confinement criterion,
$u = -\ID$ leading to well-defined charges $N^a$, can in Landau gauge be
shown by standard arguments employing Dyson-Schwinger
equations and Slavnov-Taylor identities to be 
equivalent to an infrared enhanced ghost propagator.\cite{Kug95}
In momentum space the nonperturbative ghost propagator of Landau gauge QCD  
is related to the form factor occuring in the correlations of
eq.~(\ref{Corru}) as follows,
\begin{equation}
    D_G(p) = \frac{-1}{p^2}      \, \frac{1}{ 1 + u(p^2) } \, , \;\;
                 \mbox{with}  \; \;   
                 u^{ab}(p^2)  = \delta^{ab}  u(p^2) \, . \label{DGdef}
\end{equation}
The Kugo-Ojima criterion, $u(0) = -1 $, thus entails that the Landau gauge
ghost propagator should be more singular than a massless particle pole in the
infrared. Indeed, we will present quite compelling evidence for this exact
infrared enhancement of ghosts in Landau gauge. 

The remaining dynamical aspect of confinement in this
formulation resides in the cluster decomposition property of local quantum 
field theory.\cite{Haa96} 
Including the indefinite metric spaces of covariant gauge
theories it can roughly be summarized as
follows: For the vacuum expectation values of
two (smeared local) operators $A$ and $B$, translated to a large spacelike
separaration $R$ of each other one obtains the following bounds depending on
the existence of a finite gap $M$ in the spectrum of the 
mass operator in $\mathcal{V}$,\cite{Nak90}   
\begin{eqnarray} 
        \Big|  \langle  \Omega | A(x) B(0) |\Omega \rangle  &-&         
 \langle  \Omega | A(x) |\Omega \rangle  \,  \langle  \Omega
             |  B(0) |\Omega \rangle  \Big|  \\
   && \hskip -.2cm \le  \;  \bigg\{ \begin{array}{ll} 
   \mbox{\small Const.} \, \times \, R^{-3/2 + 2N} \, e^{-MR} \!\!, \quad 
                      & \mbox{mass gap } M \; , \\
   \mbox{\small Const.} \, \times \, R^{-2 + 2N} \,, \;\; 
                      & \mbox{no mass gap} \; ,  \end{array}  \nonumber  
\end{eqnarray}
for $R^2 = - x^2 \to \infty $. Herein, positivity entails that $N = 0$, but a
positive integer $N$ is possible for the indefinite inner product structure in
$\mathcal{V}$. Therefore, in order to avoid the decomposition property
for products of unobservable operators $A$ and $B$ which 
together with the Kugo-Ojima criterion is equivalent to avoiding the
decomposition property for colored clusters, there should 
be no mass gap in the indefinite space $\mathcal{V}$. 
Of course, this implies nothing on the physical spectrum of the mass operator
in $\mathcal{H}$ which certainly should have a mass gap. 
In fact, if the cluster decomposition property holds for a product $A(x) B(0)$ 
forming a (smeared local) observable, it can be shown that both 
$A$ and $B$ are observables themselves. 
This then eliminates the possibility of scattering a physical state into
color singlet states consisting of widely separated colored clusters (the
``behind-the-moon'' problem).\cite{Nak90} 

The necessity for the absence of the massless particle pole in $\partial_\nu
F^a_{\mu\nu} $ in the Kugo-Ojima criterion shows that the (unphysical)
massless correlations to avoid the cluser decomposition property are {\em
not} the transverse gluon correlations. An infrared suppressed propagator for
the transverse gluons in Landau gauge confirms this condition. This holds
equally well for the infrared vanishing propagator obtained from
DSEs,\cite{Sme98} and conjectured in the studies of the implications of the
Gribov horizon,\cite{Gri78,Zwa92} 
as for the infrared suppressed but possibly finite ones extraced from
improved lattice actions for quite large volumes.\cite{Bon00} 

\section{The Infrared Behavior of Gluon and Ghost Propagators}
\begin{figure}[t]
 \centerline{\epsfxsize=0.9\linewidth \epsfbox{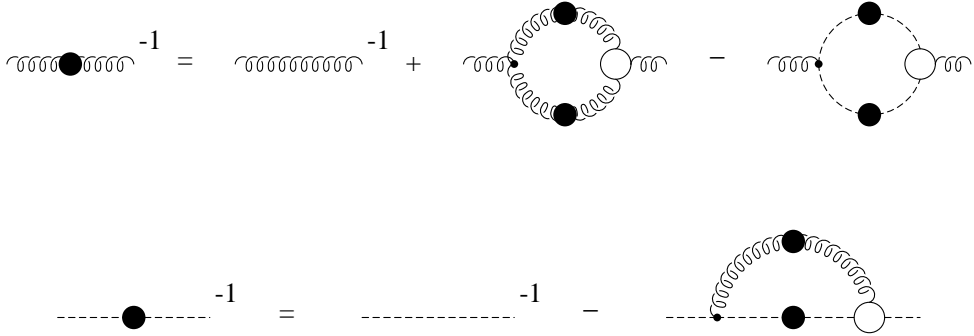}}
\caption{Diagrammatic representation of the truncated system of 
gluon and ghost DSEs.}
\label{GlGh}
\end{figure}

\begin{figure*}[t]
\parbox{.49\linewidth}{\epsfxsize=\linewidth\epsfbox{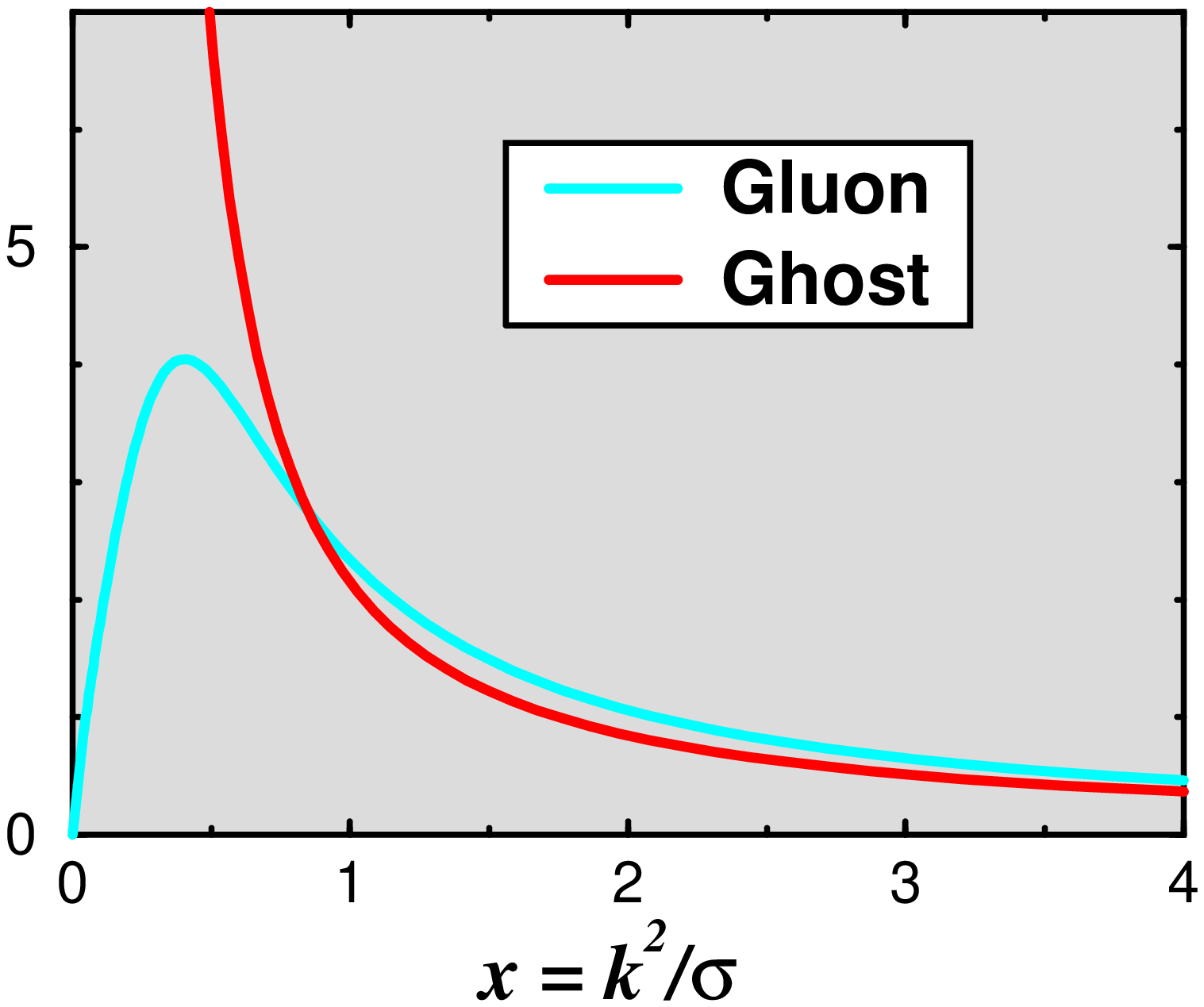}
\refstepcounter{figure} \label{ZG}
\centerline{\footnotesize Fig. \thefigure: DSE solutions for $Z(x)$ and
$G(x)$.\protect\cite{Sme98}}}  
\hskip -.2cm
\parbox{.49\linewidth}{\vskip -.5cm
\epsfxsize=1.12\linewidth\epsfbox{alpha.epsc}
\refstepcounter{figure} \label{alpha}
\vskip -.25cm 
\centerline{\hskip 1cm\footnotesize Fig. \thefigure: 
{\small $\alpha_S$} from the solutions in Fig.~\protect\ref{ZG}.}} 
\end{figure*}

In Landau gauge the two invariant functions $Z(k^2)$ and $G(k^2)$ 
in (Euclidean) momentum space parametrize the structure of the gluon and ghost
propagators, respectively, as follows (with $G(k^2) = 1/(1+u(k^2))$, {\it
c.f.}, eq.~(\ref{DGdef})),
\begin{eqnarray} 
        D_{\mu\nu}(k) = \frac{Z(k^2)}{k^2} \, \left( \delta_{\mu\nu} -
        \frac{k_\mu k_\nu}{k^2} \right)  \; ,\quad
        D_G(k)  &=& - \frac{G(k^2)}{k^2}          \;.
\end{eqnarray}
One approach that proved quite useful 
to study the non-perturbative infrared behavior of these 
functions is provided by solutions to truncated 
Dyson-\-Schwin\-ger equations(DSEs) for the propagators, {\it i.e.},
their equations of motion.\cite{Alk00,Rob00}
The known structures in the 3-point vertex functions, most importantly from
their Slavnov-Taylor identities and exchange symmtries, are thereby employed
to establish closed systems of non-linear integral equations that are
complete on the level of the gluon, ghost and quark propagators in Landau
gauge QCD. This is possible with systematically neglecting
contributions from explicit  4-point vertices to the propagator 
DSEs as well as non-trivial 4-point scattering kernels 
in the constructions of the 3-point vertices.\cite{Sme98,Alk00}  
For the pure gauge theory this leads to the propagators DSEs 
diagrammatically represented in Fig.~\ref{GlGh} with
the 3-gluon and ghost-gluon vertices (the open circles) 
expressed in terms of the two functions $Z$ and $G$. 
Employing a one-dimensional approximation one obtains the numerical 
solutions sketched in Fig.~\ref{ZG}.\cite{Sme98,Hau98} 

Asymptotic expansions of the solutions in the infrared are known 
analytically. The leading infrared behavior is thereby uniquely determined 
by one exponent $\kappa = (61 - \sqrt{1897})/19 \approx 0.92 $,
\begin{equation} 
   Z(k^2) \, \stackrel{k^2\to 0}{\sim}   
      \,  \left(\frac{k^2}{\sigma}\right)^{2\kappa}  \quad \mbox{and} \quad
          G(k^2) \, \stackrel{k^2\to 0}{\sim}    
         \, \left(\frac{\sigma}{k^2}\right)^{\kappa} \; , \label{IRB}
\end{equation}
for which the bounds $0 < \kappa < 1$ can be established.\cite{Sme98} 
The renormalization group invariant momentum scale $\sigma $ represents the 
free parameter at this point. Its relation to the scale $\Lambda$ of
perturbative QCD is rather indirect, but a rough estimate of its value may be
obtained from the corresponding running coupling shown in Fig.~\ref{alpha}
(see below). From $\alpha_S(\mu) = 0.118$ at $\mu = M_Z = 91.2$GeV, the mass
of the $Z$-boson, one then obtains $ \sigma \cong (350{\rm MeV})^2$ which, as
a qualitative test, with $M_Z/M_\tau \cong 51.5$ yields  $\alpha_S(M_\tau) =
0.38$ at the $\tau$-mass.\cite{Alk00,Sme98} 

The infrared behavior in eqs.~(\ref{IRB})
was later confirmed qualitatively by studies of further truncated
DSEs.\cite{Atk97} Neither does it thus seem to depend on the particular
3-point vertices nor on the one-dimensional approximation employed in the
original solutions. 
While the gluon propagator is found to vanish
for small spacelike momenta in this way, an apparent contradiction with
earlier DSE studies that implied its infrared enhancement\cite{Hau96} 
can be understood
from the observation that the previously neglected ghost propagator now
assumes just this: An infrared enhancement of ghosts corresponding to $u(0)
= - 1$ which alongside with the absence of massless asymptotic 
transverse gluon states for $Z(0) = 0$ is predicted by the
Kugo-Ojima confinement criterion.\cite{Kug95}  

There are also recent lattice simulations which test this criterion
directly.\cite{Nak99} Instead of $u^{ab} = -\delta^{ab}$ they obtain
numerical values of around $u = -0.7$  for the 
unrenormalized diagonal parts and zero (within  statistical errors) for the
off-diagonal parts. After renormalization, diagonal parts very close to
$-1$ result. Taking into account the finite size effects on the
lattices employed in the simulations, these preliminary results
might be perfectly reconciled with the Kugo-Ojima confinement criterion
(\ref{KO1}).

\begin{figure*}[t]
\parbox{.46\linewidth}{\hskip .5cm\epsfxsize=0.83\linewidth
\epsfbox{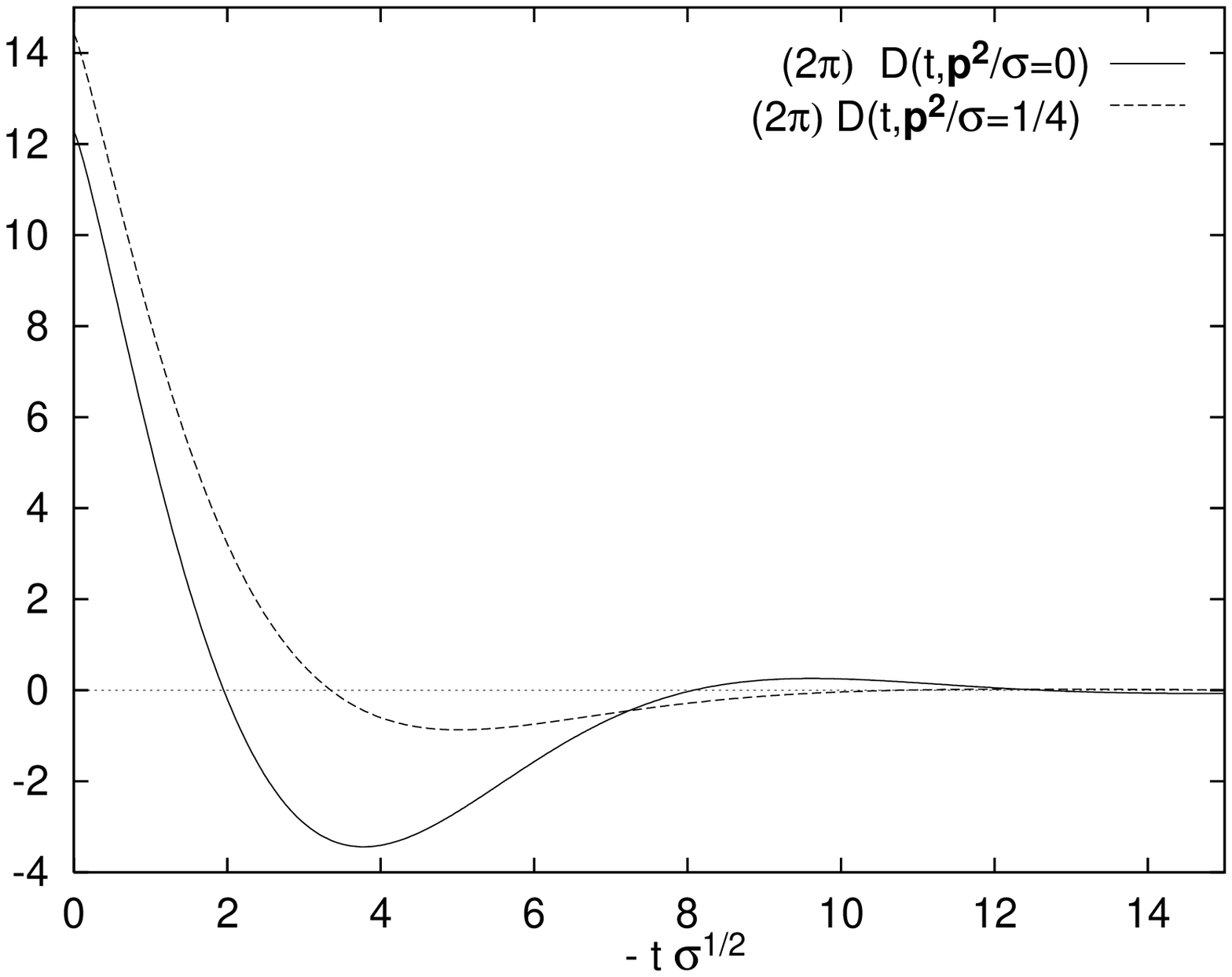}}
\hfill
\parbox{.49\linewidth}{\vskip -.7cm\hskip -.1cm
\epsfxsize=1.05\linewidth\epsfbox{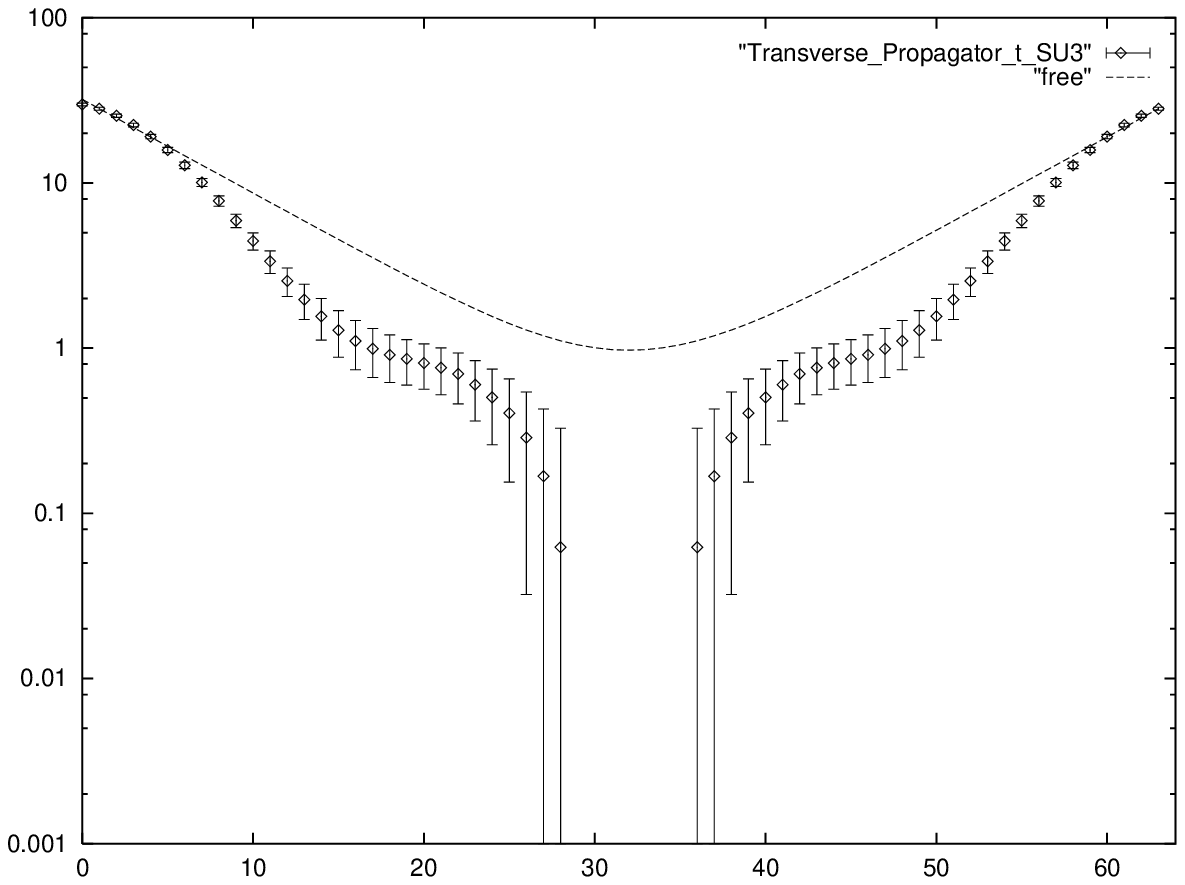}} 

\hskip .8cm\parbox{0.4\linewidth}{\refstepcounter{figure}  \label{gluon_ft}
{\footnotesize Fig. \thefigure: $D(t, \hbox{\bf p}^2)$ from DSEs for
the gluon\\[-3pt]
renormalization function $Z$ in Fig.~\ref{ZG}.}}  
\hfill
\parbox{0.4\linewidth}{\refstepcounter{figure}  \label{Nak95_Fig2}
{\footnotesize Fig. \thefigure: $D(t,\mbox{\bf p})$ at
  $(a\mbox{\bf p}) = (2\pi/48,0,0)$ for\\[-3pt] 
$\beta = 6.8$ on a $48^3   \times 64$ lattice and the free\\[-3pt] 
periodic one $\propto  \cosh(2\pi(t - N_t/2)/48)$.\cite{Nak95}}}
\vskip -.2cm
\end{figure*}

Positivity violations of transverse gluon states are manifest in 
the spectral representation of (the relevant part of) 
the gluon propagator,\footnote{This expresses the fact that 
any 2-point function is analytic in the cut complex $p^2$-plane with
singularities along the time-like real axis only which holds
due to the spectrum condition in the local description also with indefinite
metric.}  
\begin{equation} 
       D(p^2) := \frac{Z(p^2)}{p^2}  = \int_0^\infty dm^2
         \, \frac{\rho(m^2)}{p^2 +  m^2} \; .
\end{equation}
From color antiscreening and unbroken global gauge symmetry
in QCD the spectral density herein asymptotically is negative and {\em
superconvergent},\cite{Oeh80,Nis94,Alk00}  
\begin{equation} 
  \rho(k^2) \stackrel{k^2\to\infty}{\sim}  - \frac{\gamma g^2}{k^2}
  \Big(g^2 \ln\frac{k^2}{\Lambda^2}\Big)^{-\gamma-1} \hskip -.3cm ,
  \quad\mbox{and}\;
  \int_0^\infty \!\! dm^2  \rho(m^2) = \left(
\frac{g_0^2}{g^2} \right)^\gamma  \to 0 \; , 
\end{equation}
since $ \gamma > 0  $ for $ N_f\le 9 $ in Landau gauge.
This imlies that it contains contributions from quartet states 
(and does therefore not need to be gauge invariant unlike in QED).
Here, we consider the one-dimensional Fourier transform 
\begin{eqnarray}
  D(t,\hbox{\bf p}^2) =  \int \frac{dp_0}{2\pi}
    \frac{Z(p_0^2 + \hbox{\bf p}^2)}{p_0^2 + \hbox{\bf p}^2} \; e^{i p_0 t}
               \, = \, 
\int_{\sqrt{\mbox{\footnotesize\bf p}^2}}^\infty  d\omega \, \rho(\omega^2\! -\!
\mbox{\bf p}^2) \,  e^{-\omega t}\; , \label{eq:gluon_FT}
\end{eqnarray}
which for $\rho \ge 0$ had to be positive definite (and one had
$\frac{d^2}{dt^2} \ln D(t,\hbox{\bf p}) \ge 0$). 
This is clearly not the case for the DSE solution shown in
Fig.~\ref{gluon_ft} which violates reflection positivity.\cite{Alk00,Sme98} 
Even though no negative $D(t, \hbox{\bf p}^2)$ have been reported in 
lattice calculations yet, the available results,\cite{Man99} {\it e.g.}, see
Fig.~\ref{Nak95_Fig2}, agree in indicating that ln $D(t, \hbox{\bf p}^2)$ is
not the convex function of the Euclidean time it should be for positive
$\rho$.\cite{Man87,Nak95} These are non-perturbative verifications of the
positivity violation for transverse gluon states which already occur in
perturbation theory. More significant for
confinement is the fact that no massless single transverse gluon 
contribution to $\rho$ exists for $Z(0) = 0$.

In Landau gauge, the non-renormalization 
of the gluon-ghost vertex offers a convenient
possibility to define a nonperturbative running coupling.\cite{Alk00,Sme98}  
An infrared fixed point, as obtained from the DSE solutions for 
this coupling shown in Fig.~\ref{alpha},
\begin{equation}
      \alpha_{S}(\mu) =  \frac{g^2}{4\pi\beta_0} 
      Z(\mu^2) G^2(\mu^2) \stackrel{\mu \to 0}{\longrightarrow}  
 \frac{16\pi}{3N_c} \left(\frac{1}{\kappa} - \frac{1}{2} \right)^{-1} \!\!\! 
         \simeq  9.5 \; ,
\end{equation}
which determines the 2-point interactions of
color-octet quark currents, thereby 
implied the existence of the unphysical massless
states necessary to escape the cluster decomposition of colored clusters.
The corresponding massless single particle poles should occur in colored
composite operators and by virtue of the Kugo-Ojima criterion belong to
unobservable quartets. An independent verification of this result would thus
be desirable. Some implications towards an infrared finite running coupling 
as analogously being extracted from the 3-gluon vertex might be seen in lattice
calculations.\cite{Bou98} A potential problem thereby arises in presence of
the infrared enhanced ghost propagator as entailed by the Kugo-Ojima
criterion for the Landau gauge, if asymmetric momentum subtraction
schemes are employed.\cite{Alk00,Alk99} An ideal comparison would therefore 
be possible from a lattice study of the ghost-gluon vertex in a symmetric
momentum subtraction scheme.\cite{Nak99}  

\begin{figure*}[t]
\parbox{.49\linewidth}{\hskip -.5cm\epsfxsize=1.07\linewidth
\epsfbox{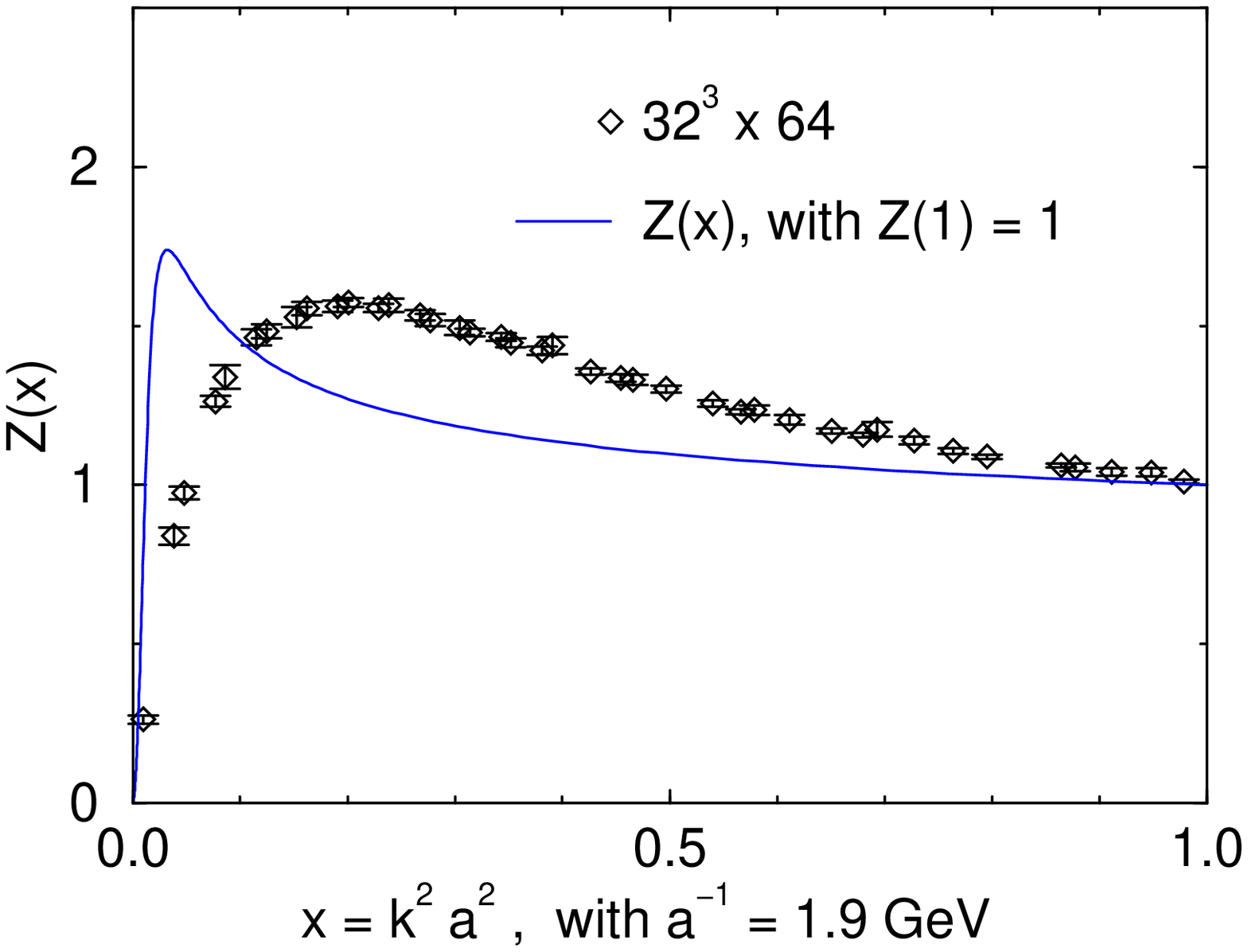}}
\hfill
\parbox{.48\linewidth}{\hskip .2cm
\epsfxsize=0.97\linewidth\epsfbox{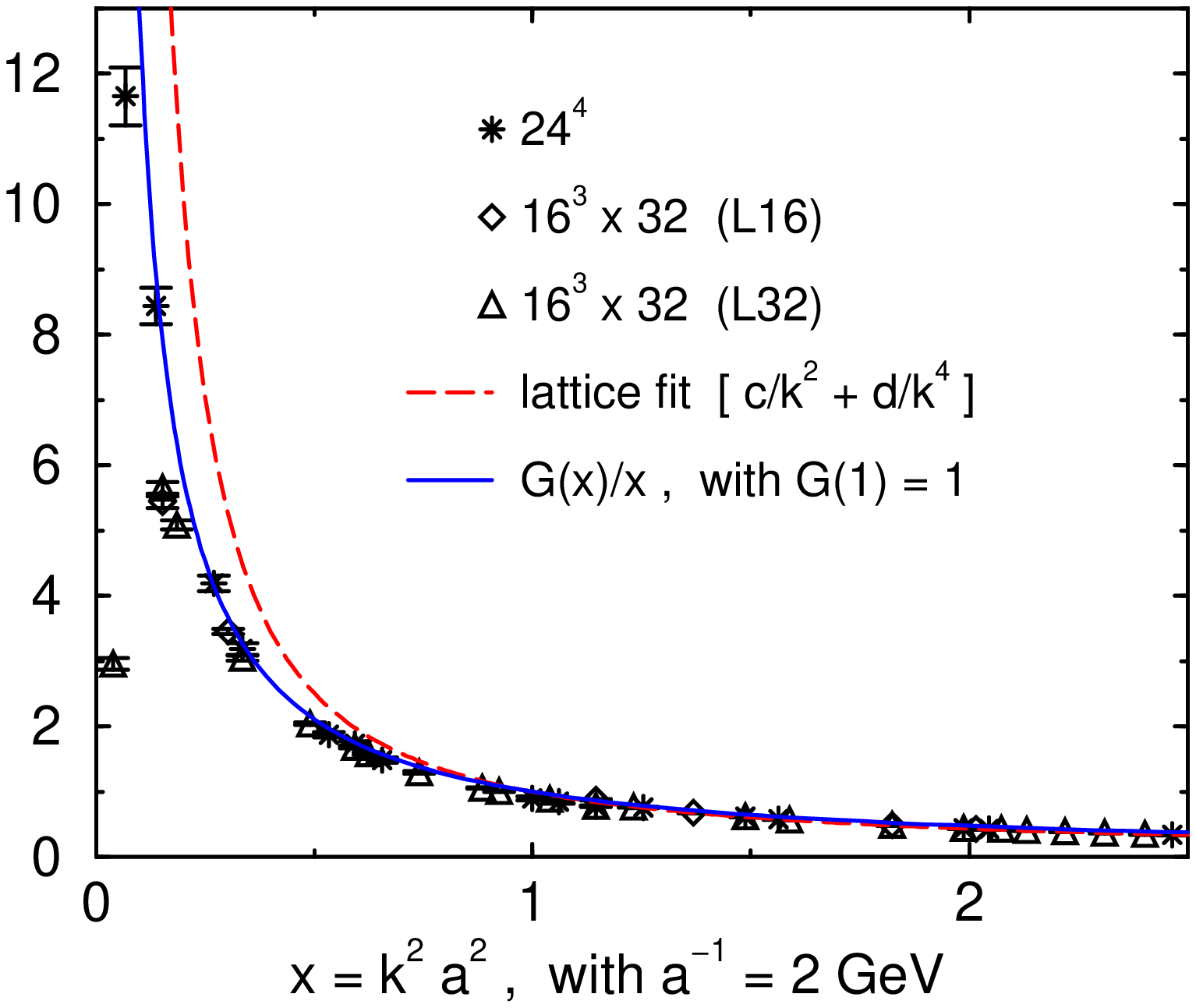}}

\hskip 0.6cm\parbox{0.4\linewidth}{
\refstepcounter{figure} \label{Der98} 
{\footnotesize Fig. \thefigure: The gluon renormalization function\\[-3pt] 
from the DSE solutions of ref.\protect\cite{Sme98} (solid line)\\[-3pt] 
and from the lattice data of ref.\protect\cite{Lei98}.}}  
\hskip 1.6cm
\parbox{0.44\linewidth}{\refstepcounter{figure}  \label{Sum95col} 
{\footnotesize Fig. \thefigure: The ghost propagator from
DSEs in ref.\protect\cite{Sme98}\\[-3pt] 
(solid line) compared to data and fit (dashed with\\[-3pt]  
$ca^2 =0.744 , \; da^4=0.256 $ for $x \ge 2$) from ref.\protect\cite{Sum96}.}}
\end{figure*}

Confirmation of the important result that the gluon renormalization function
vanishes in the infrared and no massless asymptotic transverse gluon states
occur, {\it i.e.}, $Z(0) =0$, is seen in Fig.~\ref{Der98}, where 
the DSE solution of Fig.~\ref{ZG} is compared to lattice data,\cite{Lei98} 
and it was further verified recently with improved lattice actions for large
volumes.\cite{Bon00} This infrared suppression as seen in lattice
calculations thereby seems to be weaker than the DSE result, 
apparently giving rise to an infrared finite gluon
propagator $D(k) \sim Z(k^2)/k^2 $ (corresponding to an exponent 
$\kappa = 1/2 $ in~(\ref{IRB})), but a vanishing one does not seem to be 
ruled out for the infinite volume limit.\cite{Cuc97}
Similar results with finite $D(0)$ are also 
reported from the Laplacian gauge which practically 
avoids Gribov copies.\cite{Ale00} 

The infrared enhanced DSE solution for ghost propagator is compared
to lattice data in Fig.~\ref{Sum95col}. 
One observes quite compelling agreement, the numerical DSE solution fits the
lattice data at low momenta ($x \le 1$) significantly better than the fit to an
infrared singular form with integer exponents, $D_G(k^2) = c/k^2  + d/k^4$.
Though low momenta ($x<2$) were excluded in this fit, the authors concluded
that no reasonable fit of such a form was otherwise possible.\cite{Sum96} 
Therefore, apart from the question about a possible maximum at the very
lowest momenta, the lattice calculation seems to confirm the infrared
enhanced ghost propagator with a non-integer exponent $0 < \kappa < 1$. 
The same qualitative conclusion has in fact been obtained in a 
lattice calculation of the ghost propagator in $SU(2)$,\cite{Cuc97} where its
infrared dominant part was fitted best by $D_G \sim 1/(k^2)^{1+\kappa}$ for
an exponent $\kappa $ of roughly $ 0.3$ (for $\beta = 2.7$). 

To summarize, the qualitative infrared behavior in eqs.~(\ref{IRB}), an
infrared suppression of the gluon propagator together with an infrared
enhanced ghost propagator as predicted by the Kugo-Ojima criterion for the
Landau gauge, is confirmed by the presently availabe lattice
calculations. The exponents obtained therein ($0 < \kappa \le 0.5 $) both
seem to be consistently smaller than the one obtained in solving their DSEs.
Whether the lattice data for 
the infrared behavior of both propagators  can thereby also be determined 
from one unique exponent $0<\kappa <1 $, has not yet been investigated
to our knowledge. 

L.v.S. thanks the organizers for this stimulating conference and for the
hospitality extended to him during his stay at Vienna. This work was 
supported in part by the DFG (Al 279/3-3).

\vspace{.5cm}
\hrule width 5cm

\end{document}